\documentclass[aps,showpacs,preprintnumbers,showkeys,amsmath,amssymb,superscriptaddress]{revtex4}

\usepackage{graphicx}
\usepackage{bm}
\usepackage{amsmath}
\usepackage{amsfonts}
\usepackage{color, soul}
\usepackage{float}
\usepackage{setspace}
\usepackage{multirow}
\usepackage{amssymb}
\usepackage{times}
\usepackage{cases}
\usepackage{booktabs}

\begin{document}
\title{I4/mcm-Si$_{48}$: An Ideal Topological Nodal-Line Semimetal}

\author{Laiyuan Su}
\affiliation{Hunan Key Laboratory of Micro-Nano Energy Materials and Devices, Xiangtan University, Hunan 411105, P. R. China}
\affiliation{Laboratory for Quantum Engineering and Micro-Nano Energy Technology and School of Physics and Optoelectronics, Xiangtan University, Hunan 411105, P. R. China}
\author{Shifang Li}
\affiliation{Hunan Key Laboratory of Micro-Nano Energy Materials and Devices, Xiangtan University, Hunan 411105, P. R. China}
\affiliation{Laboratory for Quantum Engineering and Micro-Nano Energy Technology and School of Physics and Optoelectronics, Xiangtan University, Hunan 411105, P. R. China}
\author{Jin Li}
\email{lijin@xtu.edu.cn}
\affiliation{Hunan Key Laboratory of Micro-Nano Energy Materials and Devices, Xiangtan University, Hunan 411105, P. R. China}
\affiliation{Laboratory for Quantum Engineering and Micro-Nano Energy Technology and School of Physics and Optoelectronics, Xiangtan University, Hunan 411105, P. R. China}
\author{Chaoyu He}
\email{hechaoyu@xtu.edu.cn}
\affiliation{Hunan Key Laboratory of Micro-Nano Energy Materials and Devices, Xiangtan University, Hunan 411105, P. R. China}
\affiliation{Laboratory for Quantum Engineering and Micro-Nano Energy Technology and School of Physics and Optoelectronics, Xiangtan University, Hunan 411105, P. R. China}
\author{Xu-Tao Zeng}
\affiliation{School of Physics, Beihang University, Beijing 100191, China}
\author{Xian-Lei Sheng}
\email{xlsheng@buaa.edu.cn}
\affiliation{School of Physics, Beihang University, Beijing 100191, China}
\author{Tao Ouyang}
\affiliation{Hunan Key Laboratory of Micro-Nano Energy Materials and Devices, Xiangtan University, Hunan 411105, P. R. China}
\affiliation{Laboratory for Quantum Engineering and Micro-Nano Energy Technology and School of Physics and Optoelectronics, Xiangtan University, Hunan 411105, P. R. China}
\author{Chunxiao Zhang  }
\affiliation{Hunan Key Laboratory of Micro-Nano Energy Materials and Devices, Xiangtan University, Hunan 411105, P. R. China}
\affiliation{Laboratory for Quantum Engineering and Micro-Nano Energy Technology and School of Physics and Optoelectronics, Xiangtan University, Hunan 411105, P. R. China}
\author{Chao Tang}
\affiliation{Hunan Key Laboratory of Micro-Nano Energy Materials and Devices, Xiangtan University, Hunan 411105, P. R. China}
\affiliation{Laboratory for Quantum Engineering and Micro-Nano Energy Technology and School of Physics and Optoelectronics, Xiangtan University, Hunan 411105, P. R. China}
\author{Jianxin Zhong}
\affiliation{Hunan Key Laboratory of Micro-Nano Energy Materials and Devices, Xiangtan University, Hunan 411105, P. R. China}
\affiliation{Laboratory for Quantum Engineering and Micro-Nano Energy Technology and School of Physics and Optoelectronics, Xiangtan University, Hunan 411105, P. R. China}

\date{\today}

\begin{abstract}
Topological semimetals (TSMs) have  attracted numerous attention due to their exotic physical properties and great application potentials. Silicon-based TMSs are of particularly importance because of their high abundance, nontoxicity and natural compatibility with current semiconductor industry. In this work, an ideal low-energy topological nodal-line semimetal (TNLSM) silicon (I4/mcm-Si$_{48}$) with a clean band crossing at Fermi level is screened from thousands of silicon allotropes by the transferable tight-binding and DFT-HSE calculations. The results of formation energy, phonon dispersion, \textit{ab initio} molecular dynamics and elastic constants show that I4/mcm-Si$_{48}$ possesses good stability and is more stable than several synthetized silicon structures. By analyzing the symmetry, it reveals that the topological nodal-line of I4/mcm-Si$_{48}$ is protected by mirror symmetry and inversion, time-reversal and SU(2) spin-rotation symmetries, and the nearly flat drumhead-like surface spectrum is observed. Furthermore, I4/mcm-Si$_{48}$ exhibits exotic photoelectric properties and the Dirac fermions with high Fermi velocity (3.4$\sim$4.36$\times$10$^5$ m/s) can be excited by low energy photons. Our study provides a promising topological nodal-line semimetal for fundamental research and potential practical applications in semiconductor-compatible high-speed photoelectric devices.
\end{abstract}


\maketitle

\section{Introduction}
In recent years, topological semimetals (TSM) have attracted numerous attention due to their exotic physical properties and great applications in the future.~\cite{WanX,PotterAC,HeLP,LiangT,XQChen,SCPMA} According to feature of the band crossing Brillouin zone (BZ), TMS can be classified into different catalogs including 0D Weyl/Dirac, triple-point and multifold fermion semimetals~\cite{WengHM,ZhuZ}, 1D nodal line semimetals~\cite{BurkovAA,Heikkila}, and 2D nodal surface semimetals~\cite{WuW,ZhangX}. For topological nodal line semimetals, band crossing in BZ can forms an extended line running across the BZ, a closed loop inside the BZ or a chain consisting of several connected loops and such band crossing can be protected by different types of symmetries~\cite{FangC}. The important signatures of a topological nodal-line semimetal is the existence of drumhead-like surface states, which is always nearly flat and can induce various exotic properties, such as high-temperature surface superconductivity~\cite{Kopnin}, strong-correlated effects~\cite{LRH}, Friedel oscillations~\cite{XQChen} and unusual transport~\cite{RWB}. Although many TNLSMs have been predicted theoretically and realized experimentally\cite{XQChen,ZhangX,LRH,YuR,Yao1,Sheng}, it is still of great interests to find ideal TNLSM candidates with clean band-crossing close to Fermi level and large characteristic dispersion energy window.~\cite{FXL}

Silicon is abundant and nontoxic and has been widely used in integrated circuits and solar cells due to its electronic properties. Besides the standard cubic-diamond silicon (CD-Si), there are also numerous three-dimensional (3D) silicon configurations or metastable forms~\cite{Wippermann}. For example, with increasing pressure up to $\sim$79 GPa, CD-Si undergoes a sequence of structural transitions to ($\beta$-Sn)-Si, Imma-Si (Si-XI), Sh-Si (Si-V), Cmca-Si (Si-VI), HCP-Si(Si-VII) and FCC-Si (Si-X). By releasing pressure, ($\beta$-Sn)-Si can  transforms into the metastable phase R8 (Si-XII) at 12 GPa and continuously to BC8 at 2GPa~\cite{Piltz}. Recently, Kim et al. have successfully synthesized a cage-like silicon crystal (Cmcm-Si$_{24}$) with open channels by a novel two-step synthesis methodology, which synthesized a Na$_4$Si$_{24}$ precursor at high pressure firstly and then removed sodium by a thermal ``degassing'' process~\cite{Kim}. L. Rapp et al. have observed ST12, BT8 and several tetragonal silicon phases by employing ultrafast laser-induced confined microexplosion, and all phases remained metastable at ambient conditions~\cite{Rapp}. There are also many other low-energy silicon semiconductors proposed by theoretical prediction~\cite{ZZS,Xiang,WangQ14,HeCYPRL} with high absorption efficiency of sunlight. P6/m-Si$_6$, proposed by Sung et al. through first-principles study at high pressure, is metallic with superconducting at 12 K and ambient pressure~\cite{Sung}. However, most of silicon allotropes are semiconductors at ambient conditions and metals under pressure~\cite{Wippermann}, the topological semimetal phases are very rare. Up to now, only two 3D silicon topological nodal-line semimetals~\cite{LiuZ} and one 2D  Dirac nodal-loop semimetal have been reported~\cite{Zhoun}. Therefore, exploring new silicon allotropes with novel properties is of crucial importance for fundamental and practical interests.

By analyzing the silicon allotropes systematically, it is found that the silicon allotropes with open channels may possess unique electronic properties different from the common silicon-semiconductors, such as TNLSMs (AHT-Si$_{24}$ and VFI-Si$_{36}$)~\cite{LiuZ} and superconducting~\cite{Sung}, and these structures may be obtained by removing Na from Na-Si clathrate~\cite{Kim}. Inspired by these experimental and theoretical works, we have screened 76 open-channel silicon structures from thousands of silicon allotropes generated by our RG2 code. Based on the electronic properties calculated by the transferable tight-binding (TB) model and first-principles calculations, an ideal silicon structure (I4/mcm-Si$_{48}$) is identified to be TNLSM with clean and closed loop band crossings at Fermi level. Symmetry analysis reveals that the nodal line in I4/mcm-Si$_{48}$ is protected by  mirror symmetry and inversion, time-reversal and SU(2) spin-rotation symmetries, resulting in nontrivial topological index and drumhead-like surface states. I4/mcm-Si$_{48}$ possesses good stability and its formation energy is lower than that of several synthesized silicon structures, thus it is highly possible that I4/mcm-Si$_{48}$ can be realized in experiments as other open-channel silicons. Especially, the Dirac fermions with high Fermi velocity can be excited by low energy photon as the strong peak in dielectric functions in low energy region, indicating its potential applications in semiconductor-compatible high-speed photoelectric devices.

\section{Method of theoretical calculations}
In this work, all the $sp^3$ silicon allotropes were generated by the graph and group theory based random strategy (RG2)\cite{HeCYPRL,ShiXZ}, and the structural and stability properties of these structures were further investigated by the Vienna ab initio simulation package (VASP)\cite{Kresse96} with the Perdew-Burke-Ernzerof generalized gradient approximation (GGA-PBE)\cite{Perdew}. The core-valence interactions were described by the PAW method\cite{PAW} with a kinetic energy cutoff of 500 eV. All atoms were fully relaxed until total energy and atomic force were less than 10$^{-6}$ eV and 0.01 eV/\AA, respectively. Monkhorst-Pack k-point meshes with a uniform density of 2$\pi$$\times$0.01\AA$^{-1}$ were generated to sample the BZ. With the optimized HSE06-based TB-parameters, all the electronic band structures of these silicon allotropes have been systematically calculated and the topological nodal-line semimetal was further confirmed by DFT-HSE calculations\cite{HSE}. The surface spectra were calculated by using the method of maximally localized Wannier functions in the WANNIER90\cite{WANNIER}.

\section{Results and discussion}
\subsection{General and transferable tight-binding model}
\begin{figure}
\includegraphics[width=\columnwidth]{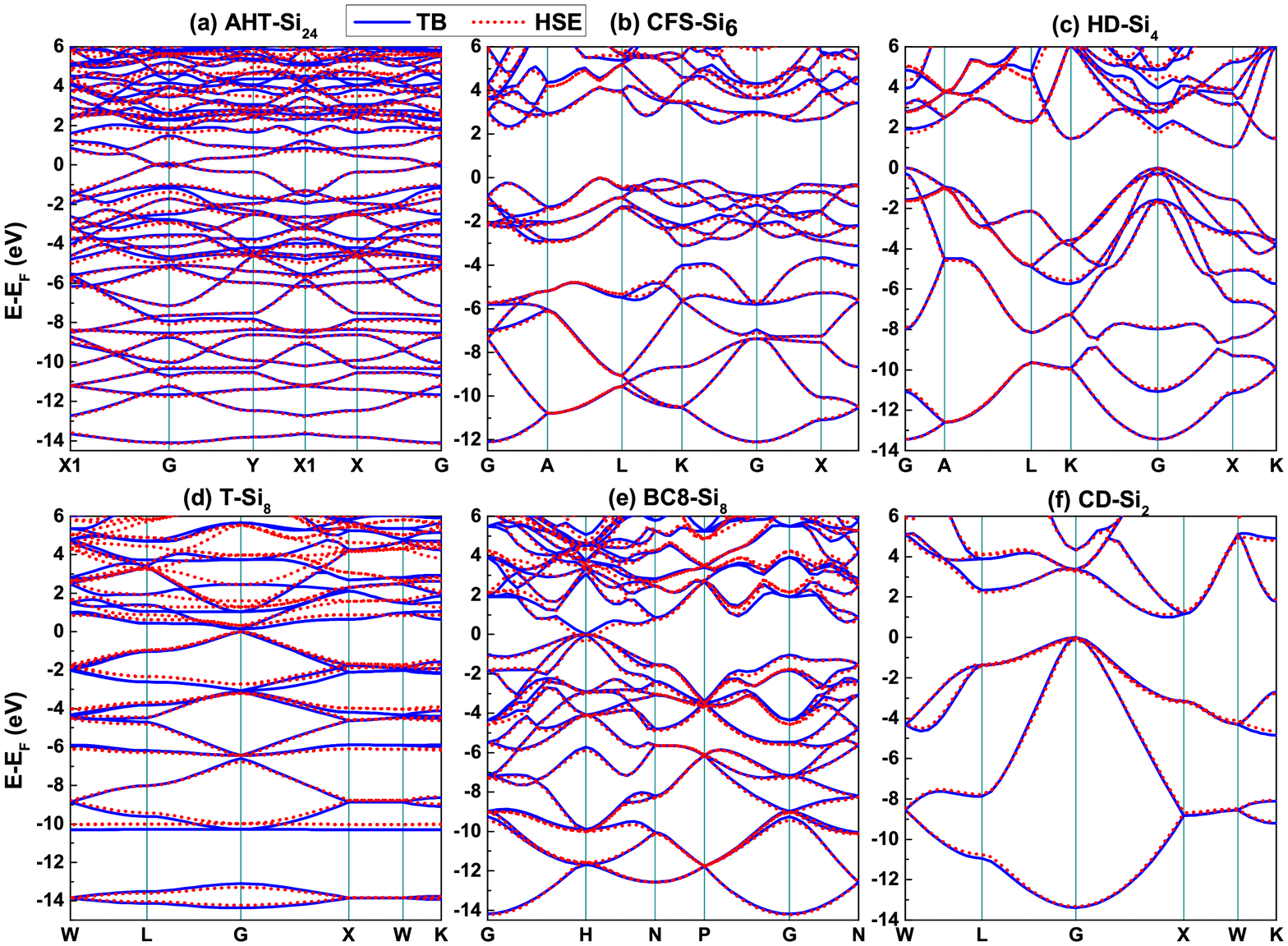}
\caption{The comparison of TB and DFT-HSE band structures of (a) AHT-Si$_{24}$, (b) CFS-Si$_6$, (c) HD-Si$_4$, (d) T-Si$_8$, (e) BC8-Si$_8$ and (f) CD-Si.}
\label{figTB}
\end{figure}

\begin{table*}[htbp]
\small
\begin{spacing}{1.5}
\caption{Tight-binding parameters for Silicon with $sp^3d^5$ basis. Units of on-site energies and hopping integrals are in eV.}
\setlength{\tabcolsep}{2.0mm}
\begin{tabular} {c|ccccccccccc}
\hline
\hline
  &  &$ss\sigma$ &$sp\sigma$ &$pp\sigma$ &$pp\pi$ &$sd\sigma$ &$pd\sigma$ &$sp\pi$  &$dd\sigma$  &$dd\pi$  &$dd\delta$  \\
\hline
\multirow{2}{0.5cm}{$t$} &$V$  &-1.821 &2.251 &1.877 &-0.126 &0.342 &1.144 &-0.367 &2.319 &0.992 &-0.060\\
                     &$q_1$  &4.124  &3.335 &2.602 &4.578  &8.577 &7.137 &7.765  &4.010 &1.959 &1.402\\
\hline
\multirow{2}{0.5cm}{$s$} &$S$  &-1.821 &2.251 &1.877 &-0.126 &0.342 &1.144 &-0.367 &2.319 &0.992 &-0.060\\
                     &$q_2$  &4.124  &3.335 &2.602 &4.578  &8.577 &7.137 &7.765  &4.010 &1.959 &1.402\\
\hline
$E_{onsite}$ & &\multicolumn{2}{c}{$E_s$=2.022} &\multicolumn{2}{c}{$E_p$=9.533} &\multicolumn{2}{c}{$E_d$=18.811}   \\
\hline
\end{tabular}
\end{spacing}
\end{table*}
It is known that the conventional DFT calculations severely underestimate the band gap, thus the hybrid functional HSE06 is always used to improve the electronic properties calculations. However, HSE06 calculations are very time consuming, especially for large unit-cell systems and high-throughput calculations with huge number of structures. To resolve this problem, we developed a general and transferable tight-binding (TB) model~\cite{SKTB,GongZH,Shixz} based on the DFT-HSE band structures. In this TB model, the $sp^3d^5$ basis is used as Si-$d$ orbitals play important roles in the TB electronic properties~\cite{NiquetYM}, and the corresponding Hamiltonian and overlap matrix could be expressed as:
\begin{equation}
 \begin{split}
&H=\sum_{i\neq i';l\neq l'}t_{il,i'l'}(c_{il}^{\dagger}c_{i'l'}+h.c.)+\varepsilon_{il}\sum_{il}c_{il}^{\dagger}c_{il}, \\
&S=\sum_{i\neq i';l\neq l'}s_{il,i'l'}(c_{il}^{\dagger}c_{i'l'}+h.c.)+\sum_{il}c_{il}^{\dagger}c_{il},
 \end{split}
\end{equation}
where $c_{il}^{\dagger}$ ($c_{il}$) and $\varepsilon_{il}$ denote the creation (annihilation) operator of an electron and the on-site energy of $l$-th orbital of $i$-th atom. $t_{il,i'l'}$ ($s_{il,i'l'}$) is the hopping (overlap) integral of an electron between the $i$-th and $i'$-th atoms. These integrals are dependent on the directional cosines of $\mathbf{R}_{ii'}$ and the bonding types in parallel and perpendicular directions to $\mathbf{R}_{ii'}$~\cite{SKTB}, and the magnitudes in each direction are assumed to scale with the atomic distance by the formula~\cite{GongZH,Shixz}:
\begin{equation}
\begin{split}
t_{il,i'l',\mu}=V_{ll'\mu}e^{q_{1}\times(1-d_{ii'}/d_{0})}, s_{il,i'l',\mu}=V_{ll'\mu}e^{q_{2}\times(1-d_{ii'}/d_{0})},
\end{split}
\end{equation}
where $d_{ii'}$ and $d_{0}$ are he distance between the $i$-th and $i'$-th atom and the reference Si-Si bond length of DC-Si (2.352 \AA), The $V_{ll'\mu}$ ($S_{ll'\mu}$) is the reference value of hopping (overlap) integral between the $l$ and $l'$ orbital with $\mu$ ($\mu=\sigma, \pi, \delta $) type bonding and $q_{1(2)}$ is the decay constants for the corresponding integrals. When $d_{ii'}$ is larger than the cutoff distance $d_{cut}$, the hopping (overlap) integral is set to be 0. More detail information about the TB parameters could be found in our previous literature~\cite{GongZH}. The above TB parameters are optimized by fitting the HSE06-based band structures of 6 small silicon structures, including AHT-Si$_{24}$, CFS-Si$_6$, HD-Si$_4$, T-Si$_8$, BC8-Si$_8$ and CD-Si. The finally optimized TB parameters are listed in the Table I and the corresponding TB band structures for these 6 selected allotropes are shown in Fig.~\ref{figTB}. It can be seen that our TB-model can describe the band structures of these six 3D silicon structures as good as the high-level HSE06 method, which reveals that out TB possesses excellent transferability for silicon allotropes with different configurations and electronic properties.

\subsection{Structure and stability of I4/mcm-Si$_{48}$}
RG2 code is a powerful tool for searching crystal structures with well-defined geometrical features. By using RG2 code, thousands initial low-density silicon allotropes containing atoms up to 60 in different symmetries have been generated and 76 new open-channel silicon structures are identified finally by geometrical checking, duplicate removing and DFT relaxation. By rapidly scanning the band structures of all the new silicon allotropes with the TB method, it is found that there are 55 semiconductors and 20 metals, while only one new structure is Dirac nodal line semimetal (I4/mcm-Si$_{48}$), which suggests that the Dirac fermion is really rare in 3D silicon structures. Therefore, we will mainly focus on I4/mcm-Si$_{48}$ in the following sections.
\begin{figure*}
\includegraphics[width=\columnwidth]{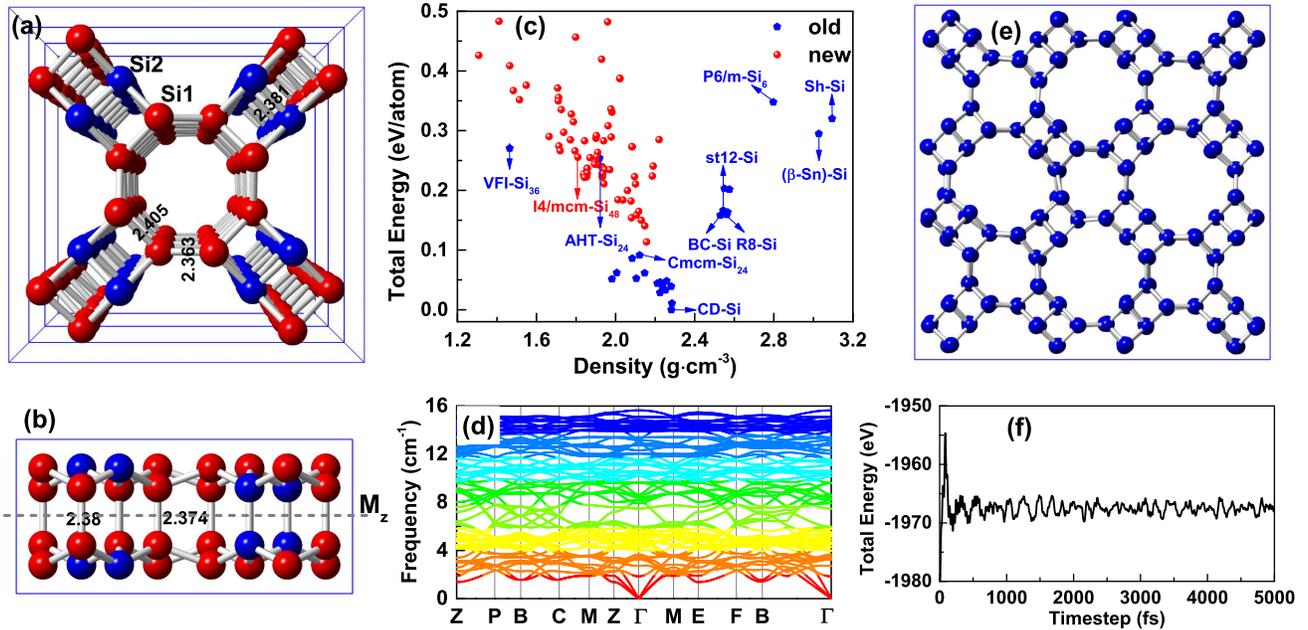}
\caption{The top (a) and side (b) views of the atomic structure of I4/mcm-Si$_{48}$ in tetragonal cell. The red and blue balls denote the two kinds of inequivalent Si atoms. (c) The calculated total energy (relative to CD-Si) as a function of the density for the old (blue pentagons) and the newly discovered (red circles) silicon allotropes. (d) Phonon dispersion of I4/mcm-Si$_{48}$ along high symmetry lines. (e) The snapshot of the atomic structure at the end of AIMD simulation at 300 K. (f) The fluctuation of total energy of I4/mcm-Si$_{48}$ during AIMD.}
\label{figstruc}
\end{figure*}

The crystal structure of I4/mcm-Si$_{48}$ is shown in Figs.~\ref{figstruc}(a, b), which contains 48 silicon atoms in its tetragonal cell and only two inequivalent atoms due to the I4/mcm (No. 140) symmetry. The relaxed lattice constants are $a=b=$13.931 \AA and $c=$ 6.378 \AA, and the two inequivalent silicon atoms locate at positions (0.0796, 0.1945, 0.1861) and (0.3068, 0.1932, 0.1866), respectively. All the Si atoms are 4-fold coordinated, and the Si-Si bond lengths and angles are in the range of 2.363 $\sim$ 2.405 {\AA} and 83.06$^\circ$ $\sim$ 120.56$^\circ$, respectively, forming distorted tetrahedrons around each Si atom. The density of I4/mcm-Si$_{48}$ is 1.81 $g/cm^3$, which is lower than that of CD-Si (2.28 $g/cm^3$), and AHT-Si24 (1.92 $g/cm^3$). As shown in Figs.~\ref{figstruc}(a), I4/mcm-Si$_{48}$ possesses the open-framwork feature with dodecagonal and hexagonal linear channels along $z$ direction, which is similar with Cmcm-Si$_{24}$\cite{Kim}, P6/m-Si$_6$\cite{Sung}, AHT-Si$_{24}$ and VFI-Si$_{36}$\cite{LiuZ}.

To estimate the stability of I4/mcm-Si$_{48}$, we firstly calculated its cohesive energy related to DC-Si and compared with several known silicon allotropes in Figs.~\ref{figstruc}(c). It is found that the cohesive energy of I4/mcm-Si$_{48}$ is about 0.255 eV higher than that of DC-Si, and it is more energetically favorable than several predicted and synthesized silicon phases, such as P/6m-Si$_6$\cite{Sung}, VFI-S$_{36}$\cite{LiuZ}, $\beta$-Sn, and SH-Si\cite{McMahon94}. The phonon dispersion of I4/mcm-Si$_{48}$ calculated by finite displacement method is shown in Figs.~\ref{figstruc}(d), and the absence of imaginary frequency strongly suggests that I4/mcm-Si$_{48}$ is dynamically stable. The thermal stability is further investigated by the ab initio MD simulations with a 2$\times$2$\times$2 supercell. The MD simulations were carried out in the canonical ensemble for 5 ps with a time step of 1 fs and the temperature was maintained at 300 K with a Nos\'{e}-Hoover thermostat. The results in Figs.~\ref{figstruc}(e, f) show that the atomic structure of I4/mcm-Si$_{48}$ remains intact with only small deformations and the total energy in AIMD oscillates within a very narrow range during the entire AIMD simulation, suggesting that I4/mcm-Si$_{48}$ is thermally stable at room temperature. In addition, the calculated elastic constants C$_{11}$, C$_{33}$, C$_{44}$, C$_{66}$, C$_{12}$ and C$_{13}$ are 83.99, 151.97, 31.96, 18.42, 49.97 and 22.30 GPa, which satisfies the mechanical stability criteria of tetragonal phase\cite{WuZJ}, i.e., C$_{ii}>$0 (i=1,2,¡, 6), (C$_{11}$-C$_{12}$)$>$0, (C$_{11}$+C$_{33}$-2C$_{13}$)$>$0 and [2(C$_{11}$+C$_{12}$)+C$_{33}$+4C$_{13}$]$>$0. These results suggest that I4/mcm-Si$_{48}$ has good stability, and it is possible to be synthesized by Na-Si clathrate as other open-channel silicon allotropes.

\subsection{Electronic properties of I4/mcm-Si$_{48}$}
\begin{figure}
\includegraphics[width=\columnwidth]{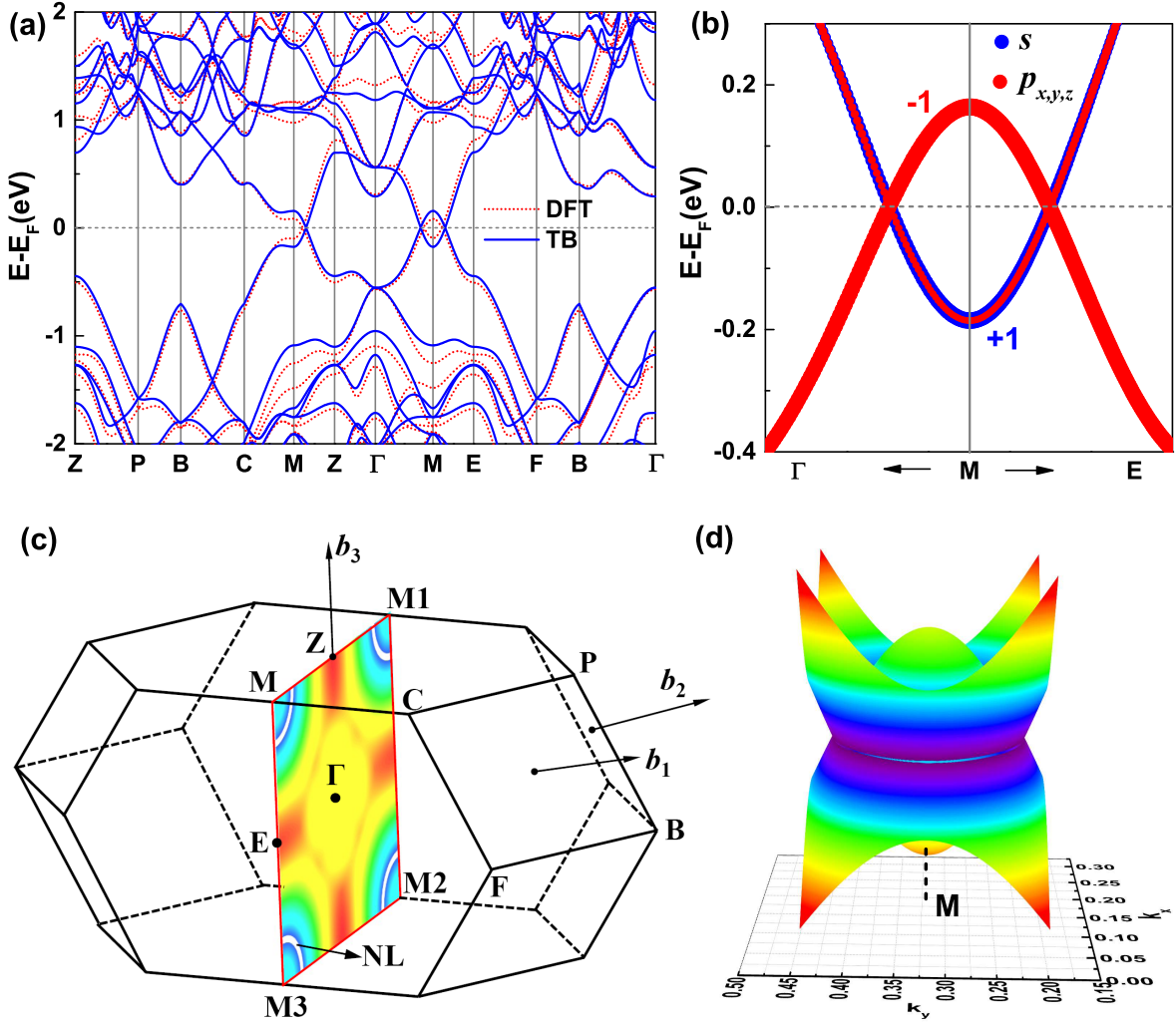}
\caption{(a) Band structure of I4/mcm-Si$_{48}$ by TB and DFT-HSE. (b) The projected TB band structure near the high symmetric k-point M. (c) The energy difference between LCB and HVB in the M-M1-M2-M3 plane. The white lines around the corners are the segments of the Dirac nodal loop. (d) 3D Dirac nodal loop formed by the valence band conduction bands near M point.}
\label{figband}
\end{figure}

The TB band structure of I4/mcm-Si$_{48}$ along the high symmetric lines in BZ is shown in Fig.~\ref{figband}(a). It can be seen that there are three Dirac-like linear crossings on the M-Z, $\Gamma$-M and M-E high symmetric lines. To investigate the origin of linear crossings, the orbital decomposed band structure around M point is calculated by TB method and plotted in Fig.~\ref{figband}(b). It is found that the highest valence band (HVB) is mainly from the $s$ orbital of Si1 atoms ($|$Si1,$s\rangle$), while the lowest conduction band (LCB) is mainly contributed by the $p$ orbital of Si2 atoms ($|$Si2,$p_{x,y,z}$$\rangle$). It is obvious that the Dirac-like linear dispersions near M are formed by the crossing of HVB and LCB due to the band inversion at M. The characteristics of I4/mcm-Si$_{48}$ band structure are further calculated by DFT-HSE calculations as shown in Fig.~\ref{figband}(a) with dash lines. One can see that the results of DFT-HSE are consistent with that of TB and the inverted band structure around M is still kept, which confirms the Dirac semimetal features of I4/mcm-Si$_{48}$.

As there are three Dirac points around M in different directions in k-space, it is necessary to investigate the distribution of these Dirac points in the entire BZ. Based on the 3D BZ shown in Fig.~\ref{figband}(c), one can see that the three Dirac points around M are in the M-M1-M2-M3 plane, and the four k-points M, M1, M2 and M3 are equivalent due to the symmetry, indicating that there are more Dirac points in this plane. Therefore, we calculated the band structure of I4/mcm-Si$_{48}$ with high density of k-points in the M-M1-M2-M3 plane by TB method. The energy difference between LCB and HVB projected in this plane is presented in the Fig.~\ref{figband}(c) and it's interesting that there are four Dirac nodal line (DNL) segments around each corner of the M-M1-M2-M3 plane as denoted by the white curves. It is expected that this four segments would form a closed nodal ring at each corner if the k-points out of the first BZ are included. Thus we calculated the 3D band structure around M as an example and it is clear that there is a Dirac nodal ring formed by the crossing of HVB and LCB as shown in Fig.~\ref{figband}(d), suggesting that I4/mcm-Si$_{48}$ is a Dirac node line semimetal. It is important that the nodal line is almost exactly on the Fermi level. For such bulk state, it was predicted that under an external magnetic field parallel to the nodal loop plane, an almost flat Landau band at the loop energy will appear, leading to a peak of density of states which can be detected by the scanning tunneling spectroscopy\cite{RhimJW}. We also calculated the Fermi velocity around the nodal line, which is a key parameter for Dirac materials, by $v_f=E(k)/\hbar|k|$ based on DFT-HSE. The $v_f$ of electrons and holes along M-$\Gamma$ (M-Z) are 3.46$\times$10$^5$ m/s (3.45$\times$10$^5$m/s) and 4.36$\times$10$^5$ m/s (4.25$\times$10$^5$ m/s), respectively, which suggests that I4/mcm-Si$_{48}$ possesses very high Fermi velocity around in Dirac-cone state and has great potential applications in high-performance electronic devices.

  \begin{figure}
    \includegraphics[width=0.75\columnwidth]{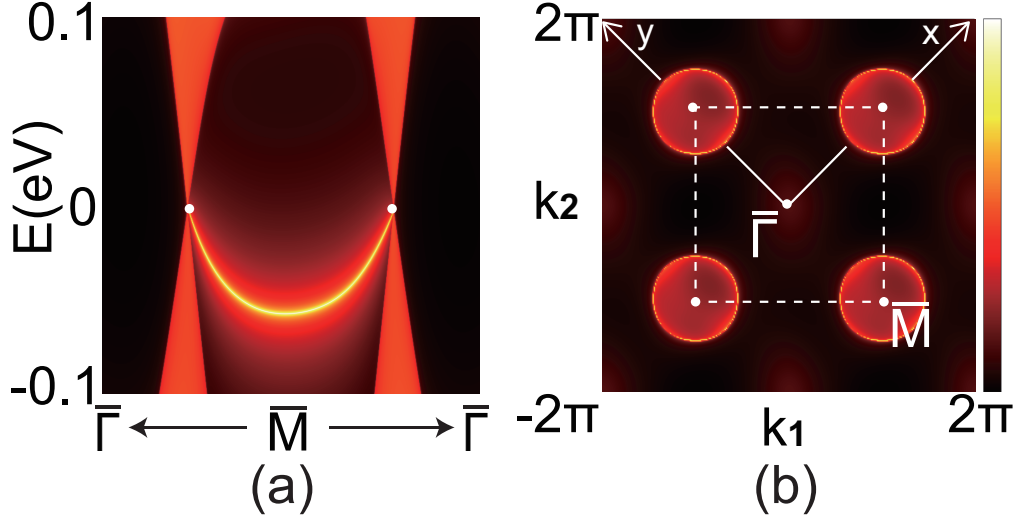}
    \caption{(a) (001) surface spectrum and (b) the corresponding Fermi loop states of I4/mcm-Si$_{48}$.}
  \label{surf140Si}
\label{figsurf}
    \end{figure}

\subsection{$k\cdot p$ model}

To characterize the low-energy band structure for the nodal-line state, we construct a $k\cdot p$ effective model. From a symmetry point of view, the little cogroup at $M$ point is $D_{4h}$ and the two crossing bands belong to two different irreducible representations $\Gamma_1^-(A_{1u})$ and $\Gamma_2^+(A_{2g})$.
Based on the generators: $\mathcal{T}=\sigma_0 \mathcal{K}$,~$\mathcal{P} = -\sigma_z$,~$\mathcal{C}_{2y} = \sigma_z$ and $\mathcal{C}_{4z} = \sigma_0$($\mathcal{C}_{2z} = \mathcal{C}_{4z}^2, \mathcal{M}_z = \mathcal{C}_{2z}\mathcal{P} $), we derive the following $k\cdot p$ effective model for the states around $M$ point:
\begin{equation}
  H_{NL}(\boldsymbol{k})= E_{0}(\boldsymbol{k})+ M(\boldsymbol{k})\sigma_z + A_1k_z\sigma_y,
  \label{EQ:NL}
  \end{equation}
where the energy and the momentum $\boldsymbol{k}$ are measured from $M$ point, $E_{0}(\boldsymbol{k})=C_0 + C_1k_z^2+C_2(k_x^2 + k_y^2)$, $M(\boldsymbol{k}) = M_0 - M_1k_z^2 - M_2(k_x^2 + k_y^2) $.
The spectrum of the effective model $H_{NL}(\boldsymbol{k})$ can be readily solved, which is given by
\begin{equation}\label{Ek}
  E_\pm({\bm k})= E_{0}(\boldsymbol{k})\pm\sqrt{M({\bm k})^{2}+A_1^{2}k_z^2}.
\end{equation}
The bands cross along a circle line with radius $\sqrt{M_0/M_2}$ around $M$ points on $k_x-k_y$ mirror-z plane. One of the most important signatures of a topological nodal-line semimetal is the existence of drumhead-like surface states. To visualize the topological surface states of I4/mcm-Si$_{48}$, we have calculated the (001) surface spectrum of the tetragonal unit cell and the corresponding Fermi loop states [See Fig.~\ref{figsurf}]. It is clear that there is a drumhead-like surface band, which connects the projected Dirac nodal points. Such topologically protected flat band was studied in several systems such as fully gapped superconductors, graphene ribbons and a nodal fermionic system\cite{RyuS,KopninNB,VolovikGE} and plays important roles for high-temperature surface superconductivity\cite{KopninNB,VolovikGE}.

The topological properties of the nodal-line can be specified by the topological invariant based on the band mirror eigenvalues. From the crystal structure shown in Fig.~\ref{figstruc}(b), one observes that I4/mcm-Si$_{48}$ preserves a $M_z$ mirror plane. The nodal line is exactly in the $\Gamma$-M-Z plane ($k_z$=0) in the first BZ, which is in $M_z$ mirror plane, as shown in Fig. 3(c). By analysing the symmetry properties of the two bands [see Fig. 3(b)], the $s$-orbital band has the $M_z$ eigenvalue +1, whereas the $p$-orbital band has $M_z$ eigenvalue $-$1. Therefore, the two bands must cross without hybridization and the nodal loop is protected by the mirror symmetry. This can be characterized by the corresponding topological invariant $\zeta_0$\cite{FangC,GaoJC}:
\begin{equation}
\begin{split}
\zeta_0=N_1-N_2,
\end{split}
\end{equation}
where $N_1$ and $N_2$ are the number of bands below the Fermi energy that has mirror eigenvalue of +1 at selected k-points on the two sides of the nodal line. As an example, we calculated the topological invariant $\zeta_0$ according to the number of bands with +1 mirror eigenvalue below Fermi energy at M and $\Gamma$ points. Our results showed that the numbers of bands with +1 mirror eigenvalue below Fermi energy at M point is $N_1=31$ and the number at $\Gamma$ point is $N_2=30$ resulting in $\zeta_0=N_1-N_2=1$, which confirms that the nodal loop is formed by the crossing of two bands with opposite mirror eigenvalues. To further investigate the robustness of TNLSM against the spin-orbital coupling (SOC) effects, the band structure of I4/mcm-Si$_{48}$ with SOC was studied. It is found that the SOC effects can be ignored completely because only very small band gap ($<$2 meV) is opened due to weak SOC of silicon[38], and the TNLSM feature is well preserved.

\begin{figure}
  \begin{center}
\includegraphics[width=1.0\columnwidth]{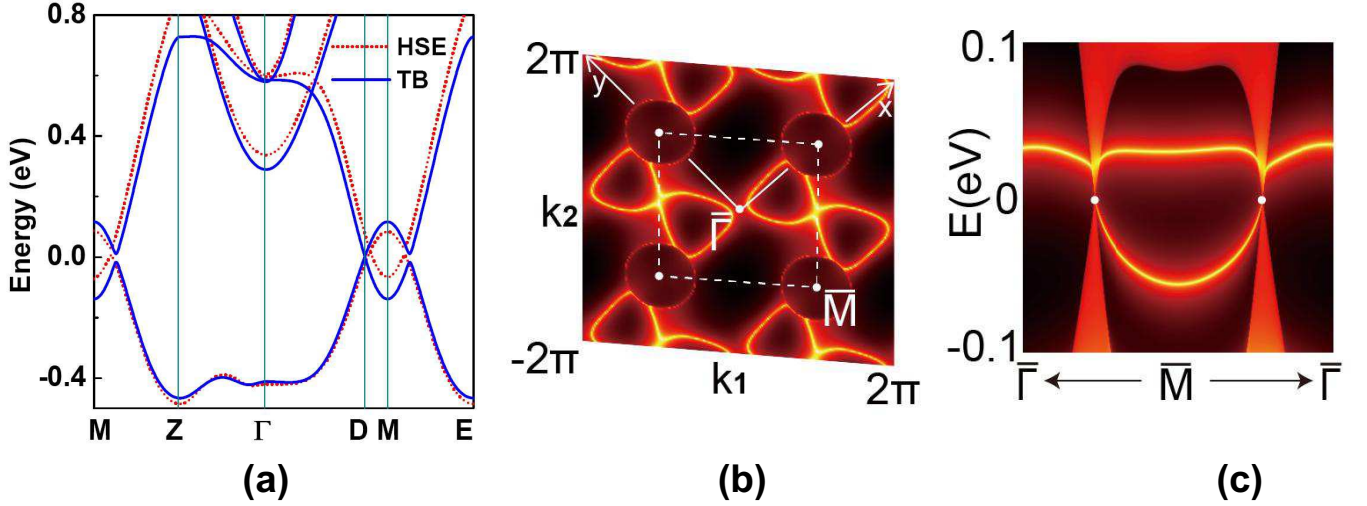}
\caption{ The band structure (a), Fermi loop states (b) and (001) surface spectrum (c) of I4/mcm-Si$_{48}$ under shear strain with $\beta=85^{\circ}$.}
\label{fig15Si}
\end{center}
\end{figure}

Besides mirror symmetry, the topological nodal-line state of I4/mcm-Si$_{48}$ is protected by inversion ($\mathcal{P}$), time-reversal ($\mathcal{T}$) and SU(2) spin-rotation symmetries, which can be demonstrated by breaking the mirror symmetry. The shear strain effects on the electronic properties of I4/mcm-Si$_{48}$ have been studied to check this point. A small strain along $x$-direction is applied by shearing $\beta$ from $90^{\circ}$ to $85^{\circ}$, which makes the symmetry of I4/mcm-Si$_{48}$ transforming from I4/mcm (No.140) to C2/C (No. 15).
The calculated band structures from DFT-HSE and TB are shown in Fig.~\ref{fig15Si}(a). One observes that the degeneracies at topological nodal-line along M-Z and M-E are removed, but the crossing point D along M-$\Gamma$ is still preserved. For this structure, the $C_{4z}$ and $M_{z}$ symmetries are broken but $\mathcal{P}$ still preserves.
The perturbation term of Hamiltonian (3) takes the form,
\begin{equation}
  \Delta H(\boldsymbol{k})=\left(C_{3} k_{x}^{2}+C_{4} k_{x} k_{z}\right) \sigma_{0}+\left(M_{3} k_{x}^{2}+M_{4} k_{x} k_{z}\right) \sigma_{z}+A_{2} k_{x} \sigma_{y}.
  \end{equation}
The spectrum of the effective model $H_{NL}^*(\boldsymbol{k}) = H_{NL}(\boldsymbol{k}) + \Delta H(\boldsymbol{k})$ is given by
\begin{equation}\label{Ek*}
  E_\pm^*({\bm k})= E_{0}^*(\boldsymbol{k})\pm\sqrt{M^*({\bm k})^{2}+(A_1k_z+A_2k_x)^2},
\end{equation}
where $E_{0}^*(\boldsymbol{k})=E_{0}(\boldsymbol{k})+C_{3} k_{x}^{2}+C_{4} k_{x} k_{z}$, $M^*(\boldsymbol{k}) =M(\boldsymbol{k}) +M_{3} k_{x}^{2}+M_{4} k_{x} k_{z} $. It is clear that band crossings occur when $M^*(\boldsymbol{k}) = 0$ and $A_1k_z + A_2k_x = 0$. By substituting $k_z = -A_2/A_1 k_x$ into $M^*(\boldsymbol{k})$, the nodal-line on $x$-$y$ plane can be described by an elliptic equation: $k_x^2/a^2 + k_y^2/b^2 = 1$, with $a = \sqrt{\frac{M_0}{M_2}}$ , and $b = A_1\sqrt{\frac{M_0}{A_1^2(M_2+M_3)-A_1A_2M_4+A_2^2M_1}}$. The shape of the nodal-line has been tilted and deformed from a circle to an oblique ellipse with its major and minor axes along $\hat{x}$ and $\hat{y}$ directions (Fig. 5(b)). And the drumhead-like surface states are found near the Fermi level as shown in Fig.5 (c).

\subsection{Optical properties of I4/mcm-Si$_{48}$}
\begin{figure}
\begin{center}
\includegraphics[width=0.6\columnwidth]{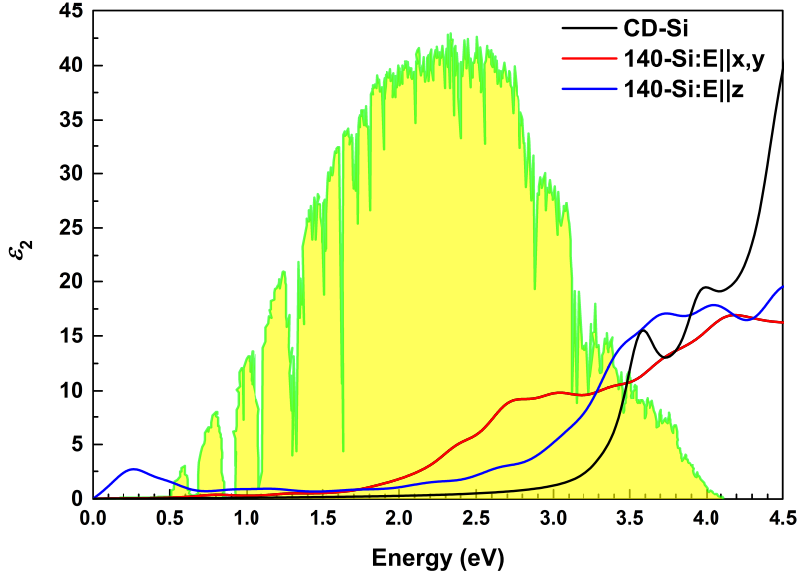}
\caption{ The imaginary part of the dielectric function $\varepsilon_2(\omega)$ of I4/mcm-Si$_{48}$ as a function of photon energy. }
\label{figoptical}
\end{center}
\end{figure}

The optical properties are of crucial importance not only for the solar cell applications of silicon-based materials but also for semimetals, such as robust edge photocurrent\cite{WangQ19}, detection of Weyl fermion chirality\cite{MaQ} and bulk photovoltaic effect (BPVE)\cite{OsterhoudtGB}. Thus, we further investigate the optical properties of I4/mcm-Si48 by the frequency-dependent dielectric function $\varepsilon(\omega)=\varepsilon_1(\omega) + i\varepsilon_2(\omega)$ by HSE-DFT calculations. The imaginary part of dielectric function $\varepsilon_2(\omega)$ is calculated by the equation\cite{Gajdos}:
\begin{equation}
 \begin{split}
\varepsilon^{(2)}_{\alpha\beta}=\frac{4\pi^2e^2}{\Omega}\lim_{q\rightarrow0}\frac{1}{q^2}\sum_{c,v,\mathbf{k}}2w_k\delta(\varepsilon_{c\mathbf{k}}-\varepsilon_{v\mathbf{k}}-\omega)\times\langle u_{c\mathbf{k}+e_\alpha \mathbf{q}}|u_{v\mathbf{k}}\rangle \langle u_{c\mathbf{k}+e_\beta \mathbf{q}}|u_{v\mathbf{k}}\rangle^*,
 \end{split}
\end{equation}
where the parameters $w_k$ is the k-point weight, $c$ and $v$ denote the conduction and valence states, respectively. $u_{c\mathbf{k}}$ is the cell periodic part of the wave-functions at $\mathbf{k}$, $\Omega$ is the unit cell volume, and $\omega$ is the photon energy. The DFT-HSE based $\varepsilon_2(\omega)$ of I4/mcm-Si$_{48}$ is shown in Fig.~\ref{figoptical} and compared to that of CD-Si with the reference air mass (AM) 1.5 solar spectral irradiance. Remarkably, I4/mcm-Si$_{48}$ exhibits stronger absorption ability than CD-Si as the significant overlap with the solar spectrum for I4/mcm-Si$_{48}$. It can be seen that I4/mcm-Si$_{48}$ shows the optical anisotropy characteristic as their optical properties in $E\parallel x(y)$ and $z$ directions are clearly different. Especially for $E\parallel z$, the optical absorption starts at 0 eV and shows distinct absorption in the low energy region from 0 to 0.7 eV. According to the band structure, it is found that the absorption in low energy region is from the interband transitions between the valence band and conduction band around the topological nodal line. These results suggest that the direct gap transitions near the nodal line are dipole allowed and the Dirac fermions with high Fermi velocity can be generated by low-energy photons, providing a new candidate for semiconductor-compatible Si-based high-speed photoelectric devices.

\section{Conclusions}
In summary, we propose a new low-energy silicon allotrope I4/mcm-Si$_{48}$ by RG2 and identify that I4/mcm-Si$_{48}$ is an ideal topological node-line semimetal with a clean band crossing loop at Fermi level due to protection of the mirror symmetry and inversion, time-reversal and SU(2) spin-rotation symmetries. I4/mcm-Si$_{48}$ possesses good stability according to the results of formation energy, phonon dispersion, \textit{ab initio} molecular dynamics and elastic constants. It is important that the nodal line is almost exactly on the Fermi level, which would produce a huge surface density of states and may lead to several interesting effects. More interestingly, the Dirac fermions with high Fermi velocity (3.4$\sim$4.36$\times$10$^5$ m/s) can be excited by low energy photon as the strong peak in dielectric functions in low energy region, which provides a promising semiconductor-compatible Si-based candidate for high-speed photoelectric devices.

\begin{acknowledgments}
This work is supported by the National Natural Science Foundation of China (Nos. 11974300, 11974299, 11704319, and 11874316), the Natural Science Foundation of Hunan Province, China (2016JJ3118 and 2019JJ50577), Scientific Research Fund of Hunan Provincial Education Department (20K127, 20A503, 20B582), Program for Changjiang Scholars and Innovative Research Team in University (IRT13093).\\
\end{acknowledgments}

\end{document}